# Integrated assessment of tritium releases from fission and fusion energy systems and their environmental implications

Liliana Arias[a], Jeffrey Wang[a], Ate Visser[b], Naofumi Akata[c], Hideki Kakiuchi[d], Shinji Ueda[d], Haruko M. Wainwright [a]

[a]*Department of Nuclear Science and Engineering, Massachusetts Institute of Technology, Cambridge, MA 02139 USA*

[b]*Nuclear and Chemical Sciences Division, Lawrence Livermore National Laboratory, Livermore, CA 94550*

[c]*Institute of Radiation Emergency and Medicine, Hirosaki University, Hirosaki, Aomori, 036-8564 Japan*

[d]*Department of Radioecology, Institute for Environmental Sciences, Rokkasho, Aomori, 039-3212 Japan*

Tritium is a naturally occurring radioactive isotope of hydrogen; a low-energy beta emitter with its health significance through internal exposure. The anticipated deployment of fusion energy systems is expected to increase tritium production significantly. Here we present an integrated assessment of tritium generation, release pathways, and resulting environmental concentrations across existing and emerging nuclear technologies. We first compile a comprehensive database of tritium production and releases from monitoring reports and literature. Analysis of light-water-reactor operating data shows that tritium discharge levels and release pathways are governed mainly by plant-specific strategies rather than reactor power or design parameters. This pattern reflects industry operating practices and efforts, with releases remaining well below regulatory limits. Our synthesis further indicates that FLiBe-based ($Li_2BeF_4$) systems may increase tritium production and mobility through enhanced permeation of tritiated hydrogen, although mitigation strategies are available to further limit emissions. Environmental transport modeling suggests that, under appropriate siting conditions, resulting tritium concentrations remain below regulatory thresholds. Furthermore, comparative risk analysis indicates that the associated carcinogenic impacts are expected to be lower than those from fossil-fuel-based power generation. Together, these findings demonstrate the importance of a comprehensive monitoring framework for contextualizing and managing routine effluent releases across energy systems.

## 1 Introduction

Fusion science and engineering have advanced significantly in recent years [1], with unprecedented levels of public support and private investment [2]. As many countries aim for the first power generation in the coming several decades, practical considerations—such as the environmental impacts of fusion energy—have gained increasing attention [3]. The majority of fusion reactor designs use the deuterium-tritium reaction to generate energy, and thus require significant amounts of tritium (on the order of tens of kilograms per $GW_{th}$ each year) as fuel [4]. Fusion reactors are expected to contain a closed-loop system for online tritium production, extraction, processing, and injection into the complex burning plasma. However, tritium ($^3H$) can permeate through metals, with permeability increasing at high temperature [5].

Consequently, several studies have suggested that fusion reactors will routinely release a significant amount of tritium to the environment [6], [7].

Tritium is a radioisotope of hydrogen with a half-life of 12.32 years [8]. It is a low-energy beta emitter, the health risk of which is internal through inhalation, absorption and ingestion similar to chemical substances [9]. Higher health risks are associated with ingesting oxidized tritium as HTO via drinking water and breathing tritiated water vapor. The short residence time of HTO in humans (~10 days) limits the exposure [10]. In contrast, the health risk of inhaling atmospheric molecular tritium (HT) is 10,000 lower [11] because absorption of HT in the lungs is very low. At the same time, animal experiments have illustrated that ingesting organically bound tritium can lead to twice the dose of comparable intake of HTO in gaseous or liquid form [12].

Tritium is naturally created in the upper atmosphere through the reactions $^{14}N(n,T)^{12}C$ and $^{16}O(n,T)^{14}N$ [13]. The annual production is estimated to be about 210 grams per year (1 g is $3.56 \times 10^{14}$ Bq) [14], while the total natural abundance of tritium is estimated to be about 3.6 kg [15]. Once created, it is highly mobile and readily incorporated into the global water cycle [9]. Global environmental tritium levels have been significantly affected by anthropogenic activities. In the past, the majority of anthropogenic tritium in the atmosphere was the result of nuclear weapons testing, particularly from thermonuclear weapons [8]. Currently, tritium is created and released by nuclear facilities such as fission power plants [13]. Different types of power plants produce and release different amounts of tritium—for example, heavy water reactors are known to release more tritium than light water reactors [16]. In addition, a significant amount of tritium is also released by spent-fuel reprocessing plants, such as the La Hague reprocessing plant in France [17].

In the past, tritium releases have caused public and regulatory concerns. As of 2024, 37 of the 54 nuclear power plant (NPP) sites in the United States have experienced leaks or spills that involved tritium concentrations greater than the drinking water standard [18], [19]. Concerns have also been expressed with regard to the discharge of tritium-containing water from the Fukushima Daiichi NPP [20], even though the amount of tritium released was lower than the routine allowed discharges from typical nuclear facilities worldwide [21].

For the successful deployment of fusion reactors, routine tritium releases must be carefully controlled to ensure that resulting concentrations in the air and water remain below regulatory limits. The environmental and public health impacts of these releases depend not only on the total production, but also on the chemical form of released tritium, release pathways and prevailing environmental conditions. In parallel, robust monitoring approaches are required, both for public safety and for nuclear security, given that tritium is also a component of nuclear weapons. Moreover, tritium releases must be evaluated and communicated within a broader context that accounts for natural background production and emissions from other facilities [17]. Although tritium production from various sources has been studied previously [18], a comprehensive synthesis of source terms across different facilities—including fusion and advanced reactors—is still lacking.

This study aims to develop an integrated and comparative assessment framework for tritium effluent releases from nuclear facilities. We first compile a source-term database—including the production, pathways, and releases—covering existing and planned facilities to enable systematic comparison across different sources and provide a broader context. We further analyze release data from currently operating facilities to identify the key factors driving variations in tritium emissions. In addition, we investigate and discuss tritium production mechanisms, release pathways, and mitigation strategies for each type of reactor. In parallel, we develop generic environmental impact models—including Gaussian plume modeling for gaseous effluents and steady-state river mixing for liquid discharges—to estimate resulting environmental concentrations. Lastly, we compare the resulting cancer risk of estimated tritium released from fusion power plants to that of chemical carcinogens released during energy production from fossil fuels. To the authors' knowledge, this is the first study to synthesize tritium generation, release pathways, management strategies, and environmental concentrations across reactor types. Despite uncertainties in future reactor releases and health impacts, this work provides a foundation for a comprehensive framework to support tritium accountability, detect anomalies and assess health and environmental impacts.

# 2 Results

## 2.1 Tritium Production and Release Database

The developed database includes tritium release data for PWR (Pressurized Water Reactor), BWR (Boiling Water Reactor), HWR (Heavy Water Reactor), HTGR (High-Temperature Gas-cooled Reactor), MSR (Molten Salt Reactor), FHR (Fluoride Salt-cooled High-Temperature Reactor), and fusion reactors (Supplementary Table 1). Release rate and energy generation values for existing BWR, PWR, HWR, and HTGR are from regulatory and other reports, while those for other reactor types are from the literature (see Methods: Tritium Release Database for the full list of references). The dataset includes measured annual releases from 62 PWRs and 31 BWRs operating commercially in the US between 2017 and 2021, and 19 HWR CANDU (Canada Deuterium Uranium) reactor units operating commercially in Canada between 2016 and 2021. Tritium release data for the HTGR were taken from the Fort St. Vrain's operation in the first half of 1986 (in addition to estimates from the literature, the values from which are consistent with the data values) [22]. Reactor release estimates for fusion plants, FHR, MSR, and HTGR are taken from six papers and reports (see Methods: Tritium Release Database).

The database allows comparison of tritium releases from different reactor types, both those existing and those planned for the future (Figure 1). The tritium releases have a large variability across the reactor types even after normalizing by the electricity production, spanning 6 orders of magnitude, ranging from a minimum of $2.61 \times 10^{-4}$ g/GWe-y for a BWR to a maximum of $8.25 \times 10^{2}$ g/GWe-y for an FHR (Figure 1). Release rates for future reactors tend to be orders-of-magnitude higher than those of existing reactors. For existing reactors, BWR had the lowest median release ($4.31 \times 10^{-3}$ g/GWe-y), followed by HTGR ($6.98 \times 10^{-2}$ g/GWe-y), PWR ($7.47 \times 10^{-2}$ g/GWe-y), and then HWR ($9.08 \times 10^{-1}$ g/GWe-y). In HWRs, $D_2O$ function as a moderator, reflector, and coolant [16], which increase tritium production compared to PWR and BWR. In contrast, HTGRs, which instead use helium gas for cooling, have lower tritium production rates [23].

For future reactors based on the literature-based estimate, fusion reactors had the lowest expected median release (2.44 g/GWe-y), followed by MSR (73 g/GWe-y), then FHR (149 g/GWe-y). We would note that the current largest release (on public record) is from the La Hague reprocessing plant (34.8 g/y, based on average yearly releases from the period 2015-2020) [24], [25], which is of similar magnitude to the total estimated yearly release from an MSR or FHR assuming 1 GWe capacity.

The higher tritium production and release expected in these reactors are largely attributed to the presence of lithium in FLiBe molten salt ($Li_2BeF_4$). FLiBe serves as a coolant in FHRs and MSRs [26], [27], and as both a coolant and breeder material in fusion reactors. In addition, MSRs dissolve the fuel in a fluoride or chloride salt, so that tritium produced as a fission product is released into the salt [28]. In addition, tritium in molten salts is primarily present as $T_2$ or TF, depending on redox potential of the solution [29]. $T_2$ can dissociate into atomic tritium and diffuse through metals before recombining on the external surface and releasing to the atmosphere [29]. $T_2$ has been shown to permeate through heat exchanger material at a rate three orders of magnitude greater than HTO under simulated ITER (a major international fusion project) conditions [29], [30]. However, most studies do not account for available mitigation strategies (Supplementary Table 5), such as nitrate-salt intermediate loops or graphite absorption, and some even assume no engineered mitigation [29]. For example, isotopic enrichment of lithium-7 (below 3 MeV, only lithium-6 produces tritium, so moderated neutrons do not produce tritium when only lithium-7 is present [31]) to levels higher than 99.9% enrichment offers a potential pathway to reduce tritium generation [26].

## 2.2 Factors Controlling Tritium Release and Pathways

For existing light water reactors (LWRs), the extensive datasets allow us to quantify the generation and release pathways in more detail, based on the actual release data. The major production is from fission of uranium and plutonium (Pu has a higher tritium yield than uranium), 1.6–2.6 g/GWe-y of tritium are produced from ternary fission [23]. This fission-derived tritium in fuel is the reason why reprocessing releases a substantial quantity of tritium from used fuel [32]. In addition, in PWRs, significant amounts of tritium are created through neutron capture of boron dissolved in the primary coolant for criticality control (about 3% of total tritium produced in a PWR: 0.073 g/GWe-y), and through neutron capture of lithium hydroxide added to the coolant for pH and corrosion control (about 0.1% of total tritium produced in a PWR: 0.002 g/GWe-y) [33], [34]. Conversely, because BWRs do not use boric acid dissolved in water for control, they rely on control rods containing boron carbide (in addition to burnable poisons and control blades) for controlling reactivity during operation [35], [36]. As such, tritium produced in BWRs (more than PWRs: 0.8 g/GWe-y) remains in the control rods, limiting its release compared to PWRs [16], [33]. The percentage of tritium released (out of the amount of tritium produced) is estimated to be 3.64% for a PWR, and 0.268% for a BWR (Figure 2).

In HWRs, the majority of tritium is formed by the $^{2}H(n,y)^{3}H$ reaction primarily in the heavy water of the moderator (about 56 g/GWe-y) and secondarily in the heat transport systems (about 0.94 g/GWe-y) [16], [37]. In addition, tritium is also formed in ternary fission in the fuel and as an activation product of boron (added to the moderator for reactivity control) and lithium (added to the Heat Transport system for corrosion control) similar to PWRs [37]. Consequently, HWRs release an order of magnitude more tritium than PWRs due to the higher production rates [37]. The percentage of produced tritium released is estimated to be 1.8% for an HWR.

In terms of the release pathways, PWRs release the majority of their tritium in liquid effluents, whereas BWRs release the majority of their tritium in gaseous effluents. BWRs and PWRs release the same order of magnitude of tritium in gaseous releases—$5.74 \times 10^{-3}$ and $7.14 \times 10^{-3}$ g/GWe-y, respectively, which is mostly associated with evaporation from spent-fuel pools [38]. PWRs release significantly more tritium in liquid effluents than BWRs ($7.12 \times 10^{-2}$ g/GWe-y for PWR, and $2.10 \times 10^{-3}$ g/GWe-y for a BWR), which is largely the filtered and treated water from the reactor coolant system/spent-fuel pool [38].

Based on the monitoring data, both gaseous and liquid tritium release from PWRs is found to be significantly correlated with total thermal energy production ($p<0.05$). The liquid release from a PWR (Figure 3a) is moderately linearly correlated with thermal energy production ($R^2$ of 0.41, p-value of $5.60 \times 10^{-8}$). The outliers for liquid tritium release from a PWR are Palo Verde's three units, which do not release liquid effluents [39]. In addition, Watts Bar Unit 1 releases more tritium than others as a result of tritium production for thermonuclear weapons [40], [41]. The gaseous release from a PWR (Figure 3c) is weakly linearly correlated with thermal energy production ($R^2$ of 0.09, p-value of $2.67 \times 10^{-2}$). Palo Verde is a high outlier for the gaseous releases as well.

On the other hand, tritium releases from BWRs were not found to significantly scale with thermal energy production for both liquid and gaseous releases ($p > 0.05$: $4.45 \times 10^{-1}$ for liquid releases and $3.48 \times 10^{-1}$ for gaseous releases) (Figure 3b) as well as total releases (Supplementary Figure 1). This is mainly attributed to the fact that some BWRs follow a policy of zero-liquid discharge, which means that large or small plants may both release very little tritium in liquid effluents [42], and the tritium inventory in spent-fuel pool is not publicly available.

Random forest regression (RFR) is then used to identify additional predictors for tritium releases. The total tritium release divided by thermal energy production is used as the target variable. The predictors are plant age, the thermal capacity, the number of loops each PWR has, the type of BWR and the containment type. The RFR performance ($R^2$ of 0.214 for PWR and -0.050 for BWR) indicates that these predictors do not adequately explain the variability in the normalized tritium releases.

### 2.3 Environmental Concentrations

The generic environmental concentration estimation tools we developed are used to calculate and visualize the concentration in air and water for different reactor types (average), assuming a 1 GWe capacity. For atmospheric pathways, different release conditions (wind speed, release height, atmospheric stability) impact the resulting air concentrations from the routine releases (Figure 4). The sources can be roughly placed into four groups, from the highest (FHR, MSR) to the lowest (PWR, BWR, HTGR). Because FHR, MSR, and fusion did not have separate liquid and gaseous effluent release information available, total releases are used for both pathways as a conservative estimate. Note that all the current fission plants have releases well below the Nuclear Regulatory Commission's (NRC) Appendix I ALARA (As Low As Reasonably Achievable) design objectives (0.1 mGy for gamma radiation or 0.2 mGy for beta radiation to individuals from gaseous effluents and 0.03 mSv/yr to the total body or 0.1 mSv/yr to any organ from liquid effluents) [43], [44].

Using the Gaussian plume model — appropriate for screening-level analysis — atmospheric tritium concentrations in the downwind direction decrease exponentially with distance from the source (after increasing at the source). Under the assumed release conditions of a wind speed of 3.4 m/s (average of U.S. windspeed data reported in [45]), neutral atmospheric stability, and a typical stack height of 100 m, all reactor releases remain at concentrations below the maximum air concentration limit to the general environment (MAC; $1 \times 10^{-14}$ g/L-air, equivalent to a total effective dose of 0.5 millisieverts from continuous inhalation over one year [46]; annual average concentration of gaseous tritium) within 7.7 km of the source — with the MSR and FHR reaching compliance at approximately 3.57 km and 7.64 km, respectively. All other reactor types remain below the MAC at every distance. For the short-duration minimum detectable concentration (MDC; $1.8 \times 10^{-16}$ g/L-air), most reactor releases are undetectable from the source, except for the three high-emission reactors: fusion power plants are detectable from about 2.49 km to 3.50 km, while MSR and FHR are detectable from about 0.768 and 0.729 km, respectively, until beyond 10 km. Under the more stringent long-duration MDC ($1 \times 10^{-17}$ g/L-air), the release from HWR, fusion, MSR, and FHR are detectable beyond 10 km, beginning at about 1.20 km, 0.886 km, 0.618 km, and 0.596 km, respectively. With respect to background tritium levels ($7.92 \times 10^{-20}$ g/L-air), all reactors exceed background levels at distances beyond 10 km from the source, except for HTGR, which remains below past ~7.67 km from the source.

Under the more dispersive conditions of a higher wind speed of 4.6 m/s (the average at the Vogtle Electric Generating Plant site [47]) combined with very unstable atmospheric stability, all reactor releases — including those from the high-emission reactors — remain below the MAC at distances greater than 810 m from the source. For the short-duration MDC, only the MSR and FHR emissions are detectable at distances approaching 2.6 km, as fusion plant emissions remain below the MDC threshold past 0.65 km. For the long-duration MDC, the two high-emission reactors remain detectable beyond 10 km, though other reactors are detectable for considerably shorter ranges than under neutral stability: the fusion power plant is detectable from ~0.20 to ~1.7 km and the HWR from~0.24 to ~1.0 km. With respect to background tritium levels ($7.92 \times 10^{-20}$ g/L-air), MSR, FHR, fusion plant, and HWR emissions exceed background concentrations beyond 10 km from the source, while the low-emission reactor releases remain below background past 1.3 km. The release estimates are very high, as one of the FHR estimates assumes no engineered mitigation [29]. Overall, these results illustrate the importance of mitigation strategies for MSR and FHR, which are already available (see Supplementary Table 5).

For the liquid release to a river, this study uses a standard International Atomic Energy Agency (IAEA) methodology for calculating river-water concentrations. The resulting concentration decreases as a function of the river discharge, assuming complete mixing in the river and no sorption and precipitation, which is typically used in the regulatory assessment (Figure 5) [48]. MSR, FHR, and fusion release curves are estimated in the case that all their tritium is released in the liquid form due to lack of data on separate gaseous and liquid effluent pathways. The highest liquid effluent tritium releases are from FHR, followed by MSR, and then fusion reactors. The lowest liquid tritium releases are from BWR, followed by HTGR as the next lowest, PWR, and then HWR.

The US river discharge ranges from first-order streams to small rivers (1.133–5.663 $m^3/s$), large rivers (70.79–283.2 $m^3/s$) and very large rivers (>283.2 $m^3/s$), according to the river classification by McManamay and DeRolph (2019) [49]. Among the rivers with the largest discharge are the Columbia River (7,500 $m^3/s$), the Ohio River (8000 $m^3/s$) and the Mississippi River (16,800 $m^3/s$). The largest sources (MSR and FHR) require the rivers with the highest class of discharge in the US (>2000 $m^3/s$) to reduce the concentrations below the regulatory drinking water limit ($2.1x10^{-12}$ g/L-water [50]), while smaller rivers are sufficient for HWR and below. Specifically, a 1 GWe FHR would require a discharge of about 3800 $m^3/s$ to be below the US tritium drinking water limit, and an MSR would require a discharge of about 2200 $m^3/s$. For the plotting range (10 $m^3/s$ to 25,000 $m^3/s$), BWR, HTGR, PWR, and HWR concentrations never exceed the US tritium drinking water limit.

The detection limit of a low-level tritium counting procedure is about $2 \times 10^{-16}$ g/L-water [51]. The background level of tritium in rivers used in this study was based on the average of the range of natural tritium levels in the Sava River and groundwater, $2 \times 10^{-15}$ g/L-water [52]. Releases from the largest sources remain above background and detectability limits at discharges greater than those of rivers in the US. HWR releases are also above detectability limits for all rivers in the US, but cross below background levels by 9,000 $m^3/s$. Releases from smaller sources cross below detectability between 350 and 13,500 $m^3/s$, and cross below background levels between 35 and 1300 $m^3/s$.

## 2.4 Cancer Risk Comparison to Fossil-Energy Systems

The health risk associated with tritium releases can be compared to carcinogen releases from fossil-energy systems, given that tritium is an internal carcinogen. The excess cancer incidence risk from inhalation at 1 km is calculated based on the same atmospheric condition and different release quantities for 1 GWe natural gas and coal power plants (Figure 6). For tritium, the cancer morbidity risk is calculated by converting the lifetime cancer risk per dose to the concentration assuming HTO vapor and 70 years of chronic inhalation [11], while lung cancer risks from chemical pollutants are taken from the published literature (Hamra et al., 2015 [53]) and agency reports such as California Office of Environmental Health Hazard Assessment (OEHHA) and U.S. Environmental Protection Agency (EPA).

The excess cancer risk associated with tritium releases from a fusion reactor is found to be almost 3 orders of magnitude lower than the lung cancer risk associated with $NO_x$ emitted by a brown coal power plant, and about 3.5 orders of magnitude lower than the lung cancer risk associated with $NO_x$ emitted by a natural gas plant (Figure 6). In addition, excess cancer risks associated with tritium are also found to be about an order of magnitude lower than cancer risks associated with persistent and non-degrading heavy metal pollutants released from a brown coal power plant. We would note that both natural gas and coal plants also release noncarcinogenic pollutants such as mercury (the full list can be found in Supplementary Table 2, 3, and 4).

# 3 Discussion

In this study, we have developed a comprehensive tritium-release database, which provides a comparative assessment of tritium releases from different nuclear facilities. Our study is motivated by previous studies that suggested that fusion reactors would release significantly more tritium than other reactor types [6], [7], [23]. Our analysis shows—based on the currently publicly available data—that fusion reactors are predicted to release about 30 times more tritium per GWe than an average PWR based on the available literature values. However, elevated tritium releases may not be unique to fusion. Several advanced fission reactors currently under development could have tritium releases comparable to fusion reactors (MSR and FHR). The large tritium releases from both advanced fission and fusion reactors are mainly attributed to large tritium production in the FLiBe coolant. Although tritium breeding in solids has been explored for fusion reactors, FLiBe is gaining popularity, particularly in high-magnetic-field fusion, due to its desirable thermal hydraulic and neutronic properties [26], [29], [54]. Although there have been studies analyzing tritium releases from currently operating reactors [55], or limited numbers of advanced fission and fusion reactors [6], [29], [56], [57], this is the first study, to the authors' knowledge, that compiles both real-world monitoring data and literature data across various reactor types.

Tritium release data from existing LWRs offers further insight on tritium management. Although tritium releases scales with thermal power for PWR, the correlation is weak and there are many outliers. Additionally, tritium releases from BWRs do not have a significant correlation with thermal power. The other quantitative plant design and operational parameters analyzed are found to not be significantly associated with tritium releases. Our results are consistent with a similar study in Korea, which found no correlation between radioactive effluent discharge and electrical output [58]. The release quantity depends more on strategies at individual plants, such as zero-liquid discharge policy in BWRs. Moreover, CANDU plants make use of tritium removal technologies to reduce environmental emissions [38], [59]. These examples illustrate that it is possible to manage and reduce tritium emissions, as well as alter the release pathways for optimizing tritium management. Such mitigation is largely attributed to the industry's self-regulating efforts based on the ALARA principle, as well as the insurance premium structure rewarding the lower tritium releases [38].

Our analysis also illustrates the importance of pathways and management such that production rates are not equivalent to release rates, and that eventual release amounts depend on a variety of factors, particularly on chemical forms of tritium. The majority of tritium production in LWRs occurs in the solid fuel, which prevents releases (note that some reactor physics codes can produce incorrect tritium inventory, since ternary fissions are ignored in some nuclear data evaluation libraries [60], [61].) Tritium production rods at Watts Bar also achieve a low permeation rate of about 0.03-0.04% [62]. Such chemical speciation is expected to be more important for the FLiBe-based fission/fusion reactors, since $T_2$ can permeate through heat exchanger material at a rate three orders of magnitude more than HTO [30]. Thus the tritium release mitigation may focus on such chemical speciation; for example, the intermediate nitrate-salt loop can oxidize tritium to HTO and/or $T_2O$ for reduced permeation (Supplementary Table 5) [26]. At the same time, because the morbidity risk coefficient for HT is 4 orders of magnitude lower than that of the morbidity risk coefficient for HTO (Federal Guidance

Report No. 13 (FGR 13) [11]), HT releases may pose a much lower health risk than HTO releases.

In parallel, our results show the importance of considering local atmospheric and hydrological conditions to evaluate the effect of tritium releases. Our atmospheric transport model results suggest that the concentrations for all facilities analyzed decrease below the regulatory limits within 10 km from the source (with releases from fusion and reactors with lower emissions never exceeding the limit), assuming an average condition. Additional dilution can also be achieved through higher release stacks, which have been demonstrated for chemical and nuclear facilities [63], [64], [65], [66]. At the same time, it is possible that FLiBe-based reactors may release tritium into effluents through permeation across primary and secondary heat exchangers [26]. Such discharge can be managed by controlling the effluent discharge rate and location. Our analysis shows that large rivers could dilute releases of even advanced nuclear reactors down to regulatory standards. Indeed, such control is already done at Watts Bar, where liquid effluent discharge is permitted only when the release from Watts Bar Dam is at least 3500 $ft^3/s$ (99 $m^3/s$) in order to have sufficient dilution [67]. Although the transport models used in this study are simplistic at the screening level, more sophisticated analysis can be developed when site-specific information and parameters are available, using models such as HYSPLIT and the AERMOD Modeling System [68], [69], [70].

At the same time, these tritium releases from nuclear facilities should be discussed in conjunction with those of carcinogens and toxic substances from other industrial facilities, given that tritium is an internal carcinogen. In the US, both the Clean Air and Water Act allow the discharge of pollutants below regulatory limits with reporting requirements [71], [72]. Our results suggest that, although advanced fission and fusion plants are expected to release more tritium than the current LWRs, their resulting carcinogenic risk is likely to be several orders of magnitude smaller than the ones from natural gas and coal power plants. Given that the tritium is the major radionuclide released from NPPs [73], [74], the expansion of fission and fusion systems, while phasing out fossil energy sources, could reduce the overall environmental carcinogenic burden from routine releases from energy systems.

We acknowledge several limitations in this study. Because our analysis is primarily data-driven, it is constrained by the availability and quality of information in the literature, particularly for future reactors. In addition, our environmental analysis is generic and at the screening-level without site specific information. However, this paper provides a screening-level synthesis across reactor types, demonstrating the value of compiling and synthesizing release data to enable comparative analysis and to provide broader perspective. Continued improvement and expansion of the database—as additional monitoring data and more accurate estimates become available—will be important for environmental protection, trend analysis, and anomaly detection, especially through encouraging commercial companies to make datasets publicly available. At the same time, although we use established methods to quantify carcinogenic risks, the different treatment of chemical and radiological carcinogens introduces uncertainties. More consistent approaches for comparing chemical and radiological risks are needed to inform energy-system choices, aiming to minimize health and environmental impacts.

# 4 Methods

## 4.1 Overview of reactor types considered in this study

The most common type of nuclear reactor in the world is a PWR, in which water is used as both a coolant and moderator (World Nuclear Association [75]). PWRs have a primary cooling circuit flowing through the reactor core under high pressure, and a secondary circuit in which steam is generated to drive the turbine (World Nuclear Association [75]). BWRs are similar to PWRs, but they have a single circuit in which water is at lower pressure than in a PWR (about 75 times atmospheric pressure), so that it boils in the core at about 285 degrees C (World Nuclear Association [75]). Steam passes through steam separators above the core and goes directly to the turbines, which need to be shielded because the water is contaminated with traces of radionuclides (World Nuclear Association [75]). HWRs are thermal reactors moderated by heavy water that are able to use natural uranium fuel due to the small thermal neutron absorption cross-section of heavy water (Han et al. [76]). CANDU reactors, a type of HWR used in Canada that this study focuses on, also use heavy water as a coolant, though light water, carbon dioxide, and organic matter can also be used as coolants (Han et al [76]). HTGRs are helium cooled, and operate at very high temperatures (coolant outlet temperature is 750°C–950°C [77]), and are graphite-moderated (Nuclear Energy Agency [78]). MSRs are reactors in which the fuel and/or coolant is in the form of a molten salt (IAEA [79]). In this study, the fuel in MSRs is a mixture of lithium and beryllium fluoride (FLiBe) salts with dissolved uranium (World Nuclear Association [80]). FHRs are a type of fission reactor that use liquid fluoride salt coolant (often FLiBe), TRISO-coated particle fuel, and graphite as the moderator (Scarlat et al. [81]). Finally, the fusion reactors analyzed in this study use FLiBe as the breeder/coolant (Pint et al. [82]). Tritium is produced when neutrons from the plasma interact with the Li in FLiBe (Pint et al. [82]).

## 4.2 Tritium Release Database

### i Release data and estimates

For light-water reactors, data on annual effluent releases is taken from NRC reports from 2017 through 2021 (NRC [83]). (Indian Point 3 is not included because it shut down in 2021). In addition, we compiled key meta datasets that are important for understanding tritium release quantities. Data on reference unit power for each nuclear reactor is taken from World Nuclear Association [84]. Thermal capacity (used to calculate thermal efficiency) for each reactor is taken from World Nuclear Association [85]. Thermal efficiency is calculated by dividing reference unit power by thermal capacity for each reactor. Data on the first grid connection for each reactor are taken from World Nuclear Association [85]. Reactor model and containment type are also taken from World Nuclear Association [85] (the containment types for South Texas 1 and Waterford 3 are not available from World Nuclear Association [85] and are instead taken from NRC [86] and NRC [87], respectively). Data on electrical generation for 2017-2021 are taken from U.S. Energy Information Administration (EIA) [88]. Thermal energy production is calculated by dividing electrical generation for each plant and year by the thermal efficiency of that plant. Data on number of loops for each PWR are taken from World Nuclear Association [85], except for the number of loops for Arkansas 1 (taken from EIA [89]), Davis Besse (EIA [90]), Millstone 2 (NRC [91]), Oconee (Nuclear Newswire [92]), Palo Verde Units 2 and 3 (EIA [93]), and St. Lucie (EIA [94]). Containment type for BWRs is taken from World Nuclear Association [85], except for Monticello (taken from NRC [95]). Data on HTGR release and thermal power capacity are taken from Public Service Company of Colorado [22]. Data on CANDU releases are taken from Canadian Nuclear Safety Commission (CNSC) [96], and data on CANDU plants' power capacity are taken from IAEA [97]. Watts Bar datapoints are only included in Figure 3, as tritium is produced for thermonuclear weapons at the Watts Bar reactors [40], [41].

Reactor release estimates for an FHR, MSR, and fusion power plant are taken from Lam et al. [29], Stempien [98], Lyu et al. [56], Cheng et al. [57], Moir [99], and Khater et al. [100]. For FHR, releases per EFPD (effective full power day) are multiplied by 365 to account for a year of release at that rate. Thermal efficiency for FHR and MSR are assumed to be 47% (U.S DOE [101]). Finally, yearly release values for La Hague, a reprocessing plant in France, are taken from Orano [24], [25].

For a fusion reactor, estimations are from multiple studies that quantify environmental releases for fusion reactors with FLiBe as the breeder/coolant (Moir [99], Khater et al. [100]). Moir [99] calculates release rate for the HYLIFE-II inertial fusion power plant design to be less than 40 Ci/day for a 1083 MWe reactor, assuming 6.5% of tritium leaks through the intermediate heat exchanger (IHX) per pass of the coolant through the IHX, and 1% leakage from the intermediate coolant through the steam generator tubes (for this study, we use the upper bound of 40 Ci/day as a conservative estimate). Additionally, Khater et al. [100] calculates routine release to be 91.5 Ci/day for OSIRIS, a 1 GWe heavy ion beam driven inertial fusion energy power reactor.

For an FHR, estimates are taken from Stempien [98] (as cited in Lam et al. [29]), which uses TRIDENT, a tritium transport code to model the 236 MWth Mk-1 FHR prototype reactor with 99.995% Lithium-7 enrichment. Estimates are taken for tritium release with no engineered mitigation, in addition to tritium releases when three different tritium-capture options were used: a permeation window, a stripping column, and a fixed graphite adsorption column.

For HTGR, both release estimates and data from a previously operating plant at Fort St. Vrain are used in order to have more datapoints for total releases, since the values for both are within an order of magnitude. For calculating the environmental concentrations, only the values from Fort St. Vrain are used, since those had separate values for gaseous and liquid effluents available (Public Service Company of Colorado [22]). For an HTGR, estimates were taken from Gainey et al. [102] for a 1000 MWth, 2000 MWth, 3000 MWth, and 4000 MWth reactor. Estimates are calculated using a simple model that accounts for (1) tritium production levels, (2) diffusion of tritium into the primary coolant from reactor materials, (3) tritium retention on reactor core materials, (4) removal of tritium by the reactor coolant purification and recovery system, and (5) tritium permeation through the steam generator tube materials, where 1-4 determine the amount of tritium in the primary coolant, and therefore the amount of tritium that can permeate through for (5). We assume thermal efficiency to be the same as the HTGR at Fort St. Vrain (39.19%).

Release estimates for an MSR are taken from Lyu et al. [56] and Cheng et al. [57]. Lyu et al. [56] calculates the tritium production in a 2 MW liquid-fueled molten salt experimental reactor (TMSR-LF1) using ORIGEN-S with an updated cross-section library generated by TRITON in SCALE 6.1.3 code system. The study assumes a 36% tritium release rate. In [57], MSR tritium release rates are calculated by summing the tritium release estimates from different structures. It assumes the tritium discharge rate is 7% in the double molten salt heat exchanger, primary pipeline, and secondary pipeline after treatment, and that the emission rate from outward filtration is 9%. Additionally, this study assumes that MSRs with different thermal power have the same tritium emission rate, which results in the tritium release values from the study being the same value when converted to g/GWe.

#### ii Tritium production

For LWRs, tritium production in different portions of the reactor are summed to create the total production rate (CNSC [16], Sabharwall et al. [23]). For a PWR, an estimate for tritium production from boric acid in the coolant is taken from Sabharwall et al. [23], production from deuterium activation and production from thermal neutron capture by lithium in the coolant is taken from the CNSC [16]. For a BWR, the estimate for tritium production from deuterium in the coolant and boron in the control rods is taken from the CNSC [16]. An estimate for the production rate of tritium as a fission product in LWRs is taken as the average of values in Sabharwall et al. [23].

Tritium production rate is taken from Sabharwall et al. [23] for HTGR. In HTGRs, the primary tritium production mechanism is through ternary fission and activation of impurities such as boron and Li-6 in the bulk graphite moderator. Tritium production rate is taken from the CNSC [16] for HWR. In HWR, the majority of tritium is produced by neutron activation of the deuterium in the heavy water moderator (Sabharwall et al. [23]).

MSR production rate is taken from Lyu et al. [56], where it is calculated with ORIGEN-S using a cross-section library generated by TRITON in SCALE 6.1.3 code system. FHR production rate is taken from Stempien [98] (as cited in Lam et al. [29]), where TRIDENT was used to model an FHR, and estimate tritium production at equilibrium for a Mk-1 FHR prototype reactor with 99.995% Li-7. For a fusion reactor, a tritium breeding ratio of ~1 is assumed, and tritium production is calculated from first principles since fusion power is directly linked to tritium production (regardless of specific fusion reactor type). Thermal efficiency is assumed to be about 45%, since conversion efficiency for future fusion plants is estimated to be between about 30 and 60% (Andlinger Center for Energy and the Environment [103]). All estimates are scaled to g/GWe-year for use in this study.

## 4.3 Statistics

Least squares linear regression (a two-tailed test) is conducted to look at the relationship between thermal energy production and tritium release for liquid and gaseous releases for both PWRs and BWRs. Outliers are defined as points with standardized residuals larger than 2.5, indicating unusually large deviation from the fitted linear regression, and fitting is based on the inliers.

## 4.4 Random Forest Regression (RFR)

Random Forest regression (RFR) is a machine-learning method developed by Breiman [104] to predict responses based on mixed numerical and categorical predictors, and to identify important predictors for given responses (Hastie et al. [105], Wainwright et al. [106]). RFR generates a large number of regression trees from bootstrapped subsampled data, and averages over all the trees (Wainwright et al. [106]). RFR is known to work well with correlated predictors similar to ridge regressions (Hastie et al. [105], Wainwright et al. [106]).

In this study, the normalized tritium release is defined as the averaged total yearly tritium release (the gas and liquid releases combined) divided by yearly thermal energy production. The normalized release for each of the LWR plants is then used as the prediction target for a RFR model using scikit-learn's RandomForestRegressor to determine the strongest predictors. The explanatory variables (predictors) used are BWR model type (*bwr_type*), containment type (*mark* for BWR and *containment* for PWR), thermal capacity (*therm_cap*), plant age (number of years since each plant was first connected to the grid (*Start))*, and number of loops for a PWR (*num_loops*). GroupKFold is used to ensure that the plants used for training were not also used for testing the model.

Plant model type and containment type explore whether different plant and containment models are associated with increased or decreased releases. Plant age explores whether accumulated inventory in the spent-fuel pool and/or plant degradation leads to increased tritium releases. Additionally, thermal capacity analyzes the relationship between power generation (thermal capacity) and tritium releases. Finally, number of loops investigates whether increased number of loops leads to increased tritium releases because of the increased surface area from which tritium can permeate.

## 4.5 Gaussian Plume Modeling

Gaussian Plume modeling is an analytical steady-state solution to the three-dimensional advection-dispersion equation, obtained by assuming constant wind velocity aligned with the positive x-axis, a wind velocity large enough that diffusion in the x–direction is much smaller than advection, isotropic diffusion, isotropic eddy diffusivities that are functions of the downwind distance only, constant emission rate from a point source, negligible topography, and no contaminant penetrating the ground (Stockie, [107]). Although it is simple, it is appropriate for a prospective analysis and generic assessment without specific site conditions (IAEA [48]).

$$C(x,y,z) = \frac{Q}{2\pi u \sigma_y \sigma_z}\left\{\exp\left(-\frac{(z-h)^2}{2\sigma_z^2}\right) + \exp\left(-\frac{(z+h)^2}{2\sigma_z^2}\right)\right\}\exp\left(-\frac{y^2}{2\sigma_y^2}\right) \tag{1}$$

where $C$ is the resulting concentration at a location *(x, y, z)* (g/m$^3$), $Q$ is the source release rate (g/s), $u$ is the wind speed (m/s), $h$ is the effective stack height (m), $y$ is the crosswind distance (m), $z$ is the vertical distance of the receptor from the ground (m), and $\sigma_Y$ (m) and $\sigma_z$ (m) are the dispersion parameters in their respective directions, determined as a function of the upwind distance $x$ and stability class (Schnadt et al. [108, p. 7], Novak and Turner [109]). In this study, we implement the Gaussian Plume model using rural parameterizations from EPA's Industrial Source Complex (ISC3) to calculate the dispersion parameters (US EPA [110]), and we assume constant wind speed and direction.

The stack height of nuclear facilities greatly varies. For example, the Calvert Cliffs Nuclear Power Plant, located in Maryland, has a stack height of roughly 60 m (NRC [111]), while the Browns Ferry Nuclear Plant in Alabama has a common stack height of about 183 m (Tennessee Valley Authority [112]). As such, we assume a stack height of 100 m in this study.

The data from the Gaussian Plume model is scaled for different types of nuclear facilities using their average yearly release rates. The concentration 1 m above ground level is plotted for distances 0 to 10 km along the downwind centerline. Calculations begin 2 m from the source to avoid results going to infinity. The wind speed of 3.4 m/s is taken from the average of US wind speed data points, in turn taken from National Laboratory of the Rockies (NLR) [45]. The wind speed of 4.6 m/s was the average wind speed measured at a height of 60 m at the Vogtle Electric Generating Plant site (NRC [47]). We check whether the resulting concentrations exceed the maximum air concentration limit of 1 x $10^{-14}$ g/L-air, equivalent to a total effective dose of 0.5 millisieverts if inhaled continuously over the course of a year [46])

Gaussian Plume modeling has various limitations. First, it does not consider stack and building wakes. Aerodynamic effects of adjacent buildings and the stack itself can alter dispersion patterns by creating turbulent zones on the sides of structures and lowering tritium plumes (CNSC [16]). If a location is known, site-specific analysis should involve HYSPLIT for resulting atmospheric concentrations, which conducts more in-depth analysis using meteorological data [68]. The AERMOD Modeling System uses meteorological data as well, and it also can account for buildings and complex terrain (US EPA [70]).

We would note that this study does not consider tritium speciation in gaseous releases (HT vs. HTO) in the environmental analysis, a factor which affects deposition and uptake. In addition, if releases are in the form of HT rather than HTO, the resulting risk is lower, as the morbidity risk coefficient for elemental tritium is 4 orders of magnitude lower than that of the morbidity risk coefficient for HTO (FGR 13 [11]).

### 4.6 Hydrological Modeling

Radionuclides discharged into surface waters experience a variety of chemical and physical processes that affect their transport from the source (IAEA [48]). These processes include flow processes (such as advection and turbulent dispersion), sediment processes, and other processes that reduce concentration in water, such as radioactive decay (IAEA [48]). A simplified and conservative approach is to use the following equation:

$$C_w, tot = \frac{Q_i}{q_r} \quad (2)$$

where $C_{w,tot}$ is the total radionuclide concentration in water ($g/m^3$), $Q_i$ is the average discharge rate for radionuclide $i$ (g/s), and $q_r$ is the mean river flow rate (defined as net freshwater velocity times river width times flow depth) ($m^3/s$) (IAEA [48]). We assume that decay is negligible compared to the transport time in typical rivers, since tritium has a half-life of 12.32 years (Tsipis [8]). We convert $C_{w,tot}$ to units of g/L-water by dividing by 1000.

In this study, we assume that there is complete mixing with the receptor point farther than seven times the river depth [48]. We also assume continuous radionuclide discharge and constant river flow rate. Additionally, we assume the river initially contains no tritium concentration upstream. We use pre-calculated concentrations based on the unit release and multiply it by the different release rates.

Higher river discharge leads to lower tritium concentrations. No dilution before release is assumed, although this is a method that can lead to lower environmental concentrations. The river concentration analysis makes a number of simplifying assumptions, and so further hydrological analysis should involve hydrology codes such as Amanzi-ATS and ParFlow, which can be used for surface water modeling.

### 4.7 Web Interface Development

An interactive webpage is created in which users can calculate resulting atmospheric concentration. The application is built on the Shiny for Python framework, and is deployed as a static web application on GitHub Pages. Python execution is enabled via WebAssembly (shinylive / Pyodide). Users can input wind speed in miles per hour, atmospheric stability (A-F), and release height (stack height, m). The default stack height is 100 m. They can also choose from the following facilities to plot air releases: PWR, BWR, HWR, MSR, a fusion power plant, HTGR, or FHR. They also have the option to add a custom gas release curve. In addition, in a second tab, users can input a minimum and maximum discharge in $m^3/s$. Users can also choose to plot liquid releases in a river from the aforementioned facilities.

### 4.8 Cancer Risk Comparison to Fossil-Energy Systems

#### i Atmospheric transport

To obtain resulting risk of cancer incidences at 1 km (Figure 6), a power plant size of 1 GWe is assumed. The concentration at 1 km from the source is obtained using Gaussian Plume modeling assuming 3.4 m/s windspeed (average US windspeed from NLR [45]) and neutral atmospheric stability. We assume release from a 150 m stack, roughly the stack height of coal plants, for fair comparison between different energy systems. (The mean stack height of coal-fired power plants with available stack height data in the U.S. EPA's 2020 National Emission Inventory was found to be 155 m (Kim et al. [113])). The resulting concentration value is multiplied by the excess cancer incident risk per concentration for each pollutant to find the resulting excess cancer incident risk for each pollutant.

#### ii Radiation Excess Risk

For radionuclides including tritium, cancer risk can be interpreted as the average risk per unit exposure for a lifetime in the US EPA Federal Guidance Report No. 13, where the risk coefficients apply to populations that approximate the age, gender, and mortality experience characterized by the 1989-91 U.S. decennial life tables (FGR 13 [11]). In order to evaluate cancer risk from inhalation, this paper uses the morbidity coefficient (defined as an estimate of average total risk of experiencing a radiogenic cancer) for tritium from US EPA FGR No. 13 [11].

The most important forms of tritium (from an atmospheric behavior perspective) in atmospheric plumes from nuclear facilities are HTO and HT, while tritiated methane accounts for only a small amount of atmospheric tritium (CNSC [16]). Because the morbidity risk coefficient for elemental tritium is 4 orders of magnitude lower than that of the morbidity risk coefficient for HTO, we use the risk coefficient for HTO to be conservative (FGR 13 [11]). The HTO risk coefficient is a gender-averaged, age-averaged value. Although only a small fraction of HT is converted to HTO in the atmosphere (likely less than 0.4%) (Kim et al. [114]), the dominant mechanism for the formation of atmospheric HTO from released HT is through interaction with the soil (Brown et al. [115]).

Using the long-term inhalation rates in US EPA [116], the average amount of air inhaled in a lifetime of 70 years is estimated to be 377 million liters. (Lifetime is often assumed to be 70 years for exposure assessment (US EPA [117])). We converted the cancer risk per amount tritium to the excess cancer risk per atmospheric concentration, using the average amount of air inhaled in a lifetime of 70 years and assuming the chronic life exposure to be consistent with chemical pollutants.

#### iii Chemical Excess Risk

We compare the resulting excess risk from chemical pollutants released during electricity production from coal and natural gas to that of tritium releases from a fusion plant. First, emission factors and their 95% confidence intervals for public electricity and heat production (in units of mass/GJ) are taken for natural gas, hard coal, and brown coal from the European Environment Agency/ European Monitoring and Evaluation Programme Air Pollutant Emission Inventory Guidebook 2023 [118]. The factors are converted to units of g/GWe-y assuming thermal efficiency of 35% (U.S. Department of Energy (DOE) [119]) for natural gas plants and 32.5% for both coal power plants (Congressional Research Service, [120]) (Supplementary Table 2, 3, and 4). It is important to note that some of the factors are derived from measurement data below the limit of quantification (marked by “<”), thereby representing upper limits rather than precise values, but they can still be used directly for calculation (European Environment Agency/ European Monitoring and Evaluation Programme [118]). We note that for electricity production from coal, the TSP, $PM_{10}$ and $PM_{2.5}$ emission factors represent filterable PM emissions.

For chemical pollutants, inhalation unit risk (IUR) is defined as an estimate of the increased cancer risk from chronic inhalation exposure to a concentration of 1 μg/m$^3$, and can be multiplied by lifetime exposure in μg/m$^3$ to estimate the lifetime cancer risk (EPA [121]). We look at increased cancer risk due to release of heavy metal pollutants and $NO_x$. For arsenic, the IUR from OEHHA [122] is used. For cadmium, the IUR from EPA [123] is used. The IUR for Ni is taken from OEHHA [124], and the IUR for Pb is taken from OEHHA [125]. Other heavy metals without total IUR available were not included when calculating the resulting risk of cancer incidences (such as Cr).

The IUR of the different pollutants is calculated in the various sources through use of epidemiological data from both human and animal studies through a variety of different methods. For example, the inhalation unit risk for cadmium is calculated using a two-stage extrapolation method (EPA [123]). For nickel and nickel compounds, data from occupational exposure of Ontario refinery workers is used to calculate inhalation unit risk from cumulative exposure and SMR (standardized mortality ratio) values (OEHHA [126]). A relative risk model is used for linear extrapolation to low dose lifetime exposure for the data (OEHHA [126]). Exposure is adjusted for equivalent lifetime exposure, and excess relative risk for lifetime exposure is estimated considering the background lifetime mortality risk for Ontario (OEHHA [126]). The upper bound of risk in (μg/m$^3$)$^{-1}$ is used as the inhalation unit risk (OEHHA [126]). For lead, due to inadequate availability of human data, inhalation unit risk is calculated using data from a study in which rats were given lead acetate in their feed (OEHHA [126]). Doses are converted to human equivalent doses and using the GLOBAL86 computer software, a linearized multistage model is fit to the male kidney tumor dose-response data to yield a maximum likelihood estimate of the slope term, which relates the probability of cancer to the carcinogen dose (OEHHA [126]). Since available human data suggests that about 50% of inhaled lead is absorbed compared to about 10% of ingested lead (summarized by Owen [127]), assuming that the percentage is similar for humans and rats, and that an average adult human has a body weight of 70 kg and air intake of 20 m$^3$, the results from the rat study are applied to humans, and the upper confidence limit is used as the inhalation unit risk (OEHHA [126]). Overall, based on limited data availability and varying methods used for calculating inhalation unit risk, there exists significant uncertainty in the comparison.

For $NO_x$, we estimate IUR using available epidemiological data, since IUR values were not available in the government agency reports. We do not include $PM_{2.5}$ since the risk is relatively low compared to $NO_x$ and for electricity production from coal, $PM_{2.5}$ emission factors represent filterable PM emissions. Relative risk of lung cancer for $NO_x$ is taken from Hamra et al., 2015 [53], while baseline lung cancer incidence rate for the United States is taken from the National Cancer Institute to be 5.2% of people over their lifetimes (based on 2021–2023 data) [128]. In Hamra et al., 2015 [53], for a 10 μg/m$^3$ increase in $NO_x$, lung cancer incidence rate increases by 3% (95% confidence interval ranges from 1% to 5%). We can then represent relative risk, *RR*, using the equation

$$RR(C) = e^{\beta \Delta C} \tag{3}$$

Where $\Delta C$ is the concentration increase in μg/m$^3$, and $\beta$ is the relative growth rate in units of m$^3$/μg. The equation can be rearranged to solve for $\beta$.

In addition, we can thus represent the excess risk, $R$, in units of cases per person over a lifetime, by the equation

$$R(C) = R_0\left(e^{\beta \Delta C} - 1\right) \tag{4}$$

Where $R_0$ is the baseline cancer risk, $\Delta C$ is the concentration increase due to the electricity production in μg/m$^3$, and $\beta$ is the relative growth rate in units of m$^3$/μg (so that the excess risk is 0 when the exposure concentration is equivalent to the baseline concentration). The IUR can thus be estimated by dividing the resulting excess lifetime risk $R$ by the excess pollutant concentration, $\Delta C$. The 95% confidence interval for excess lifetime risk at 1 km is found using the 95% confidence interval values for emissions and the 95% confidence interval values for the increased cancer risk due to increased $NO_x$ concentration.

# 5 Figures

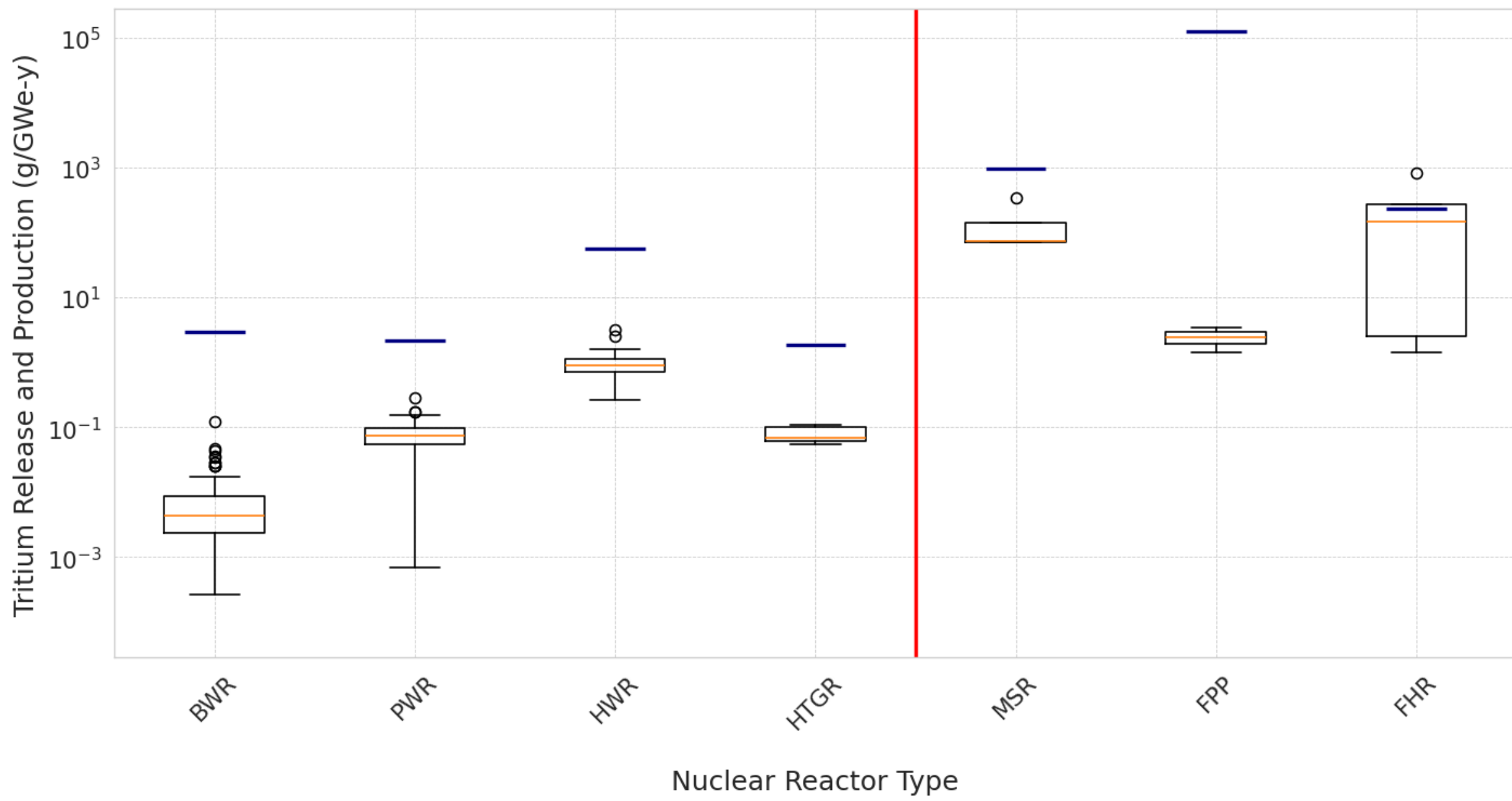

*Figure 1: Boxplots of release rates in different types of nuclear reactors. The blue lines represent theoretical production rates for each reactor, while the orange lines represent the median release rate for each reactor. The red line separates reactors for which some*

*actual data was used from reactors where only literature values were used. The box itself ranges from the 25th percentile to the 75th percentile (interquartile range) release rate for each type of reactor. The whiskers represent 1.5 times the interquartile range from the box. The circular points represent release-rate outliers for each reactor. In the case of an FHR, the estimate for yearly tritium production was lower than the highest estimate for tritium releases. This is because the production value used is for operation at equilibrium, and more tritium can be released when more is produced at beginning of life. For HTGR, we mixed actual data and literature values, since the values are consistent.*

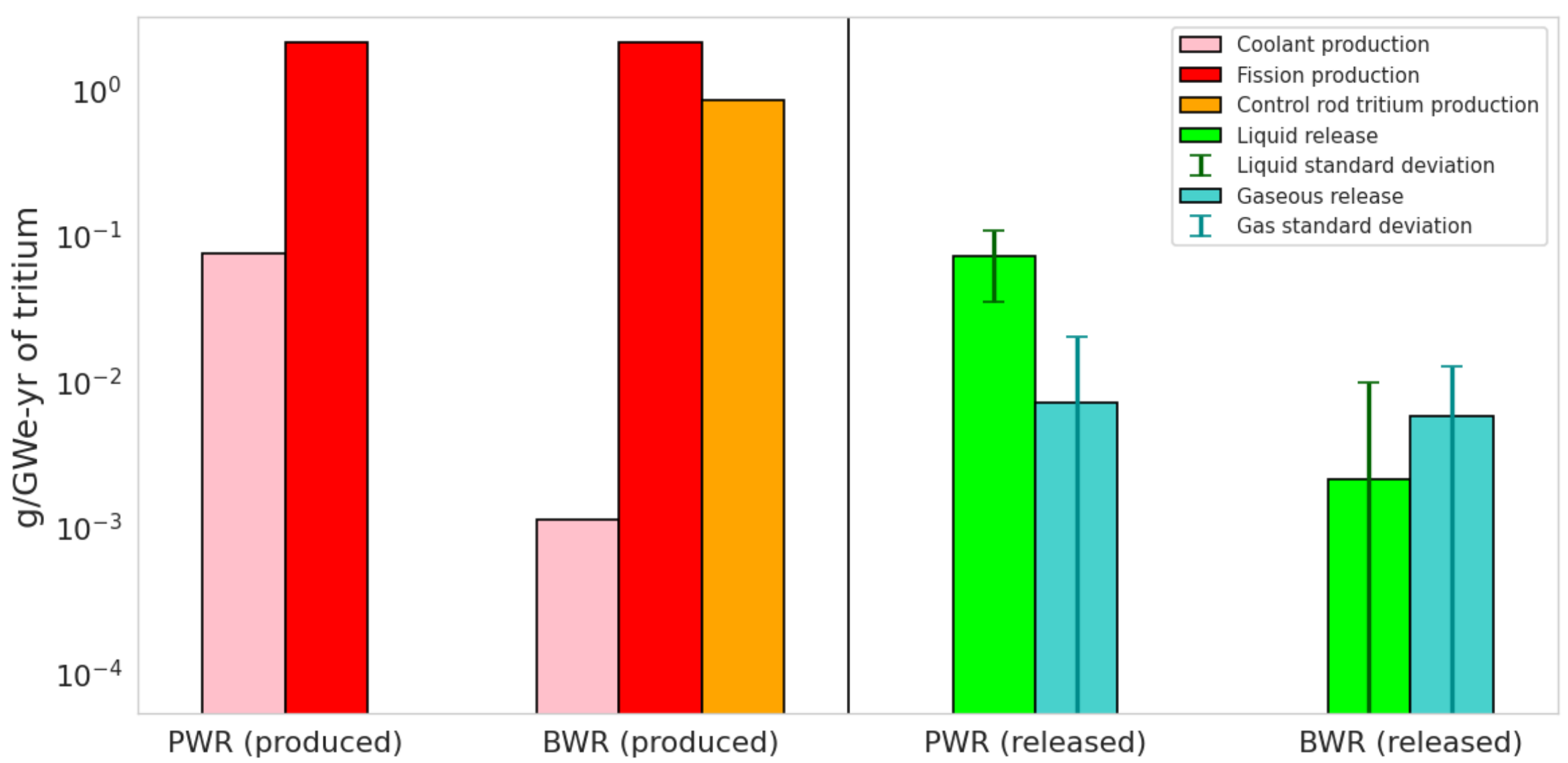

*Figure 2: Tritium production vs. environmental release in g/yr for a 1 GWe plant, plotted on a log scale. Estimates from the literature for tritium production in different parts of the reactor (fuel, coolant, control rods) are from [16], [23]. In addition, empirical routine release values are plotted as the means of the data, normalized for a 1 GWe plant. Error bars represent 1 standard deviation from the mean.*

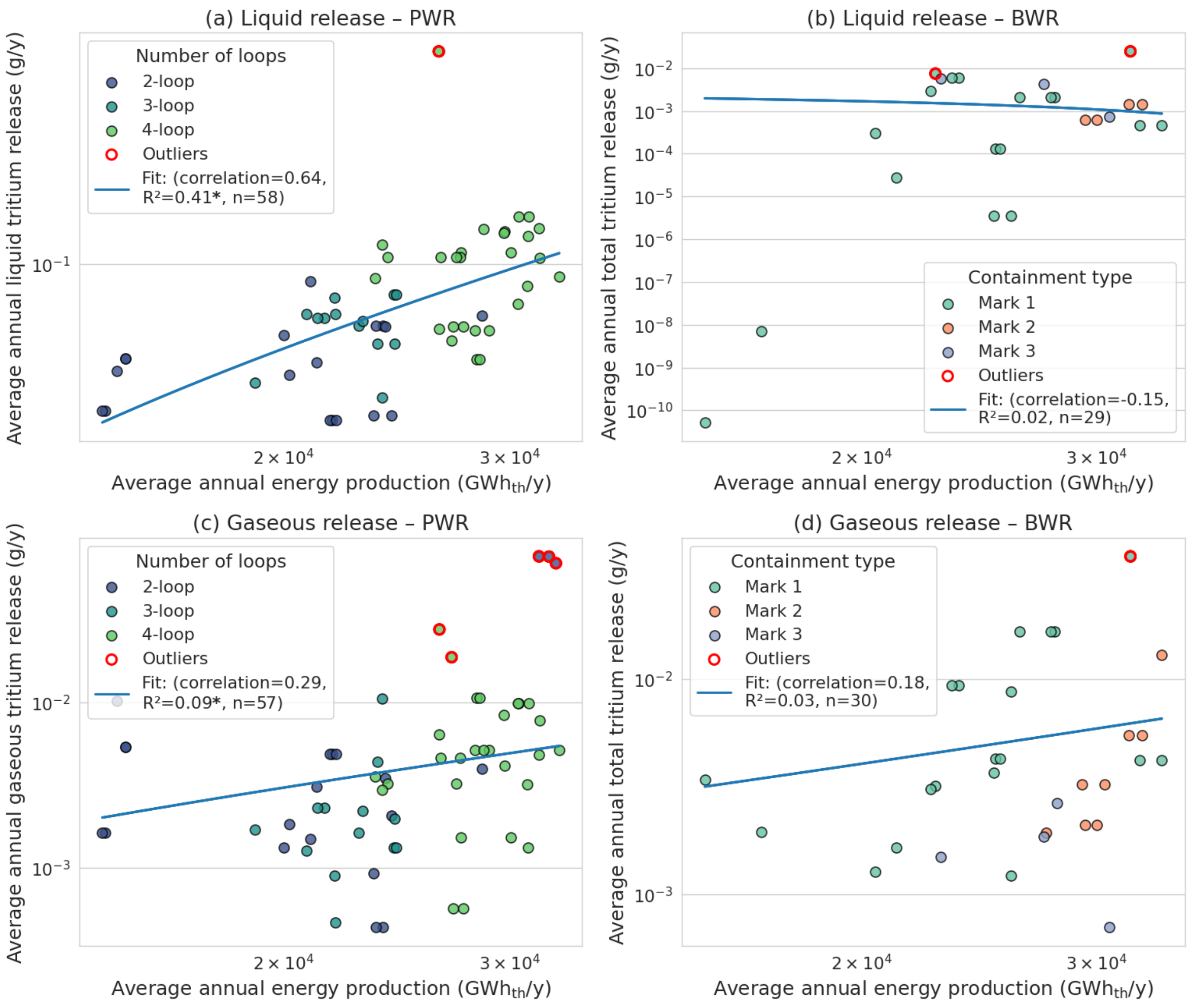

*Figure 3: Relationship between thermal annual energy production and tritium release for LWRs. Different colored points correspond to different numbers of loops for PWRs, and different types of containment for BWRs. Linear regression was conducted to look at the relationship between thermal energy production and tritium release for liquid and gaseous releases for both PWRs and BWRs. Outliers were defined as points with standardized residuals larger than 2.5, indicating unusually large deviation from the fitted linear regression. Fitting is based on the inliers. An asterisk next to the $R^2$ value indicates the slope of the regression is statistically significant ($p<0.05$).*

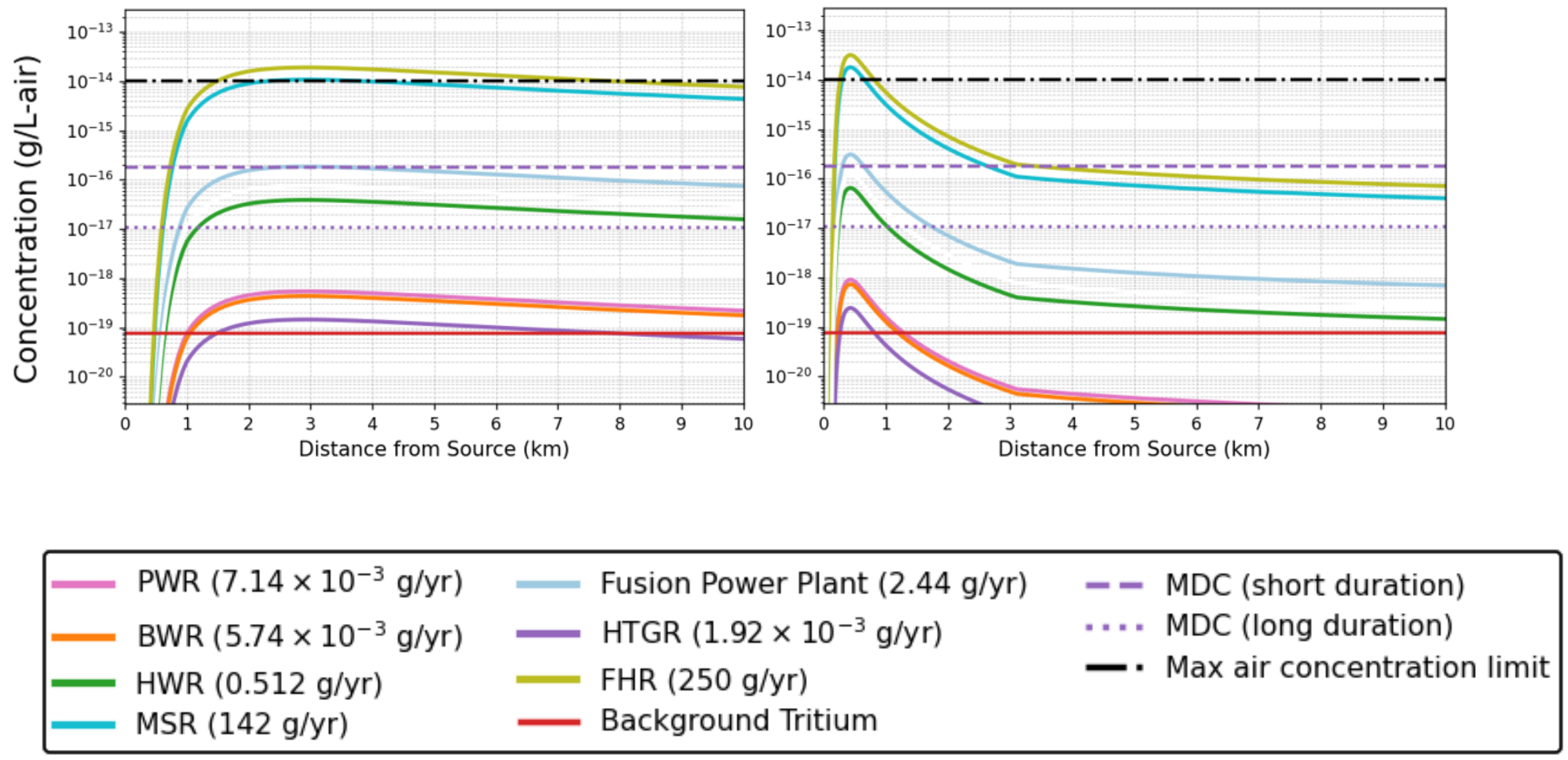

*Figure 4: Resulting atmospheric concentration of tritium 1 m above the ground as a function of distance along the downwind centerline for different nuclear facilities (all plants 1 GWe) with a stack release height of 100 m, a wind speed of (a) 3.4 m/s (average of US wind speed data points taken from [45]) and neutral atmospheric stability and (b) 4.6 m/s (average wind speed measured at a height of 60 m at the Vogtle Electric Generating Plant site [47]) and very unstable atmospheric stability.*

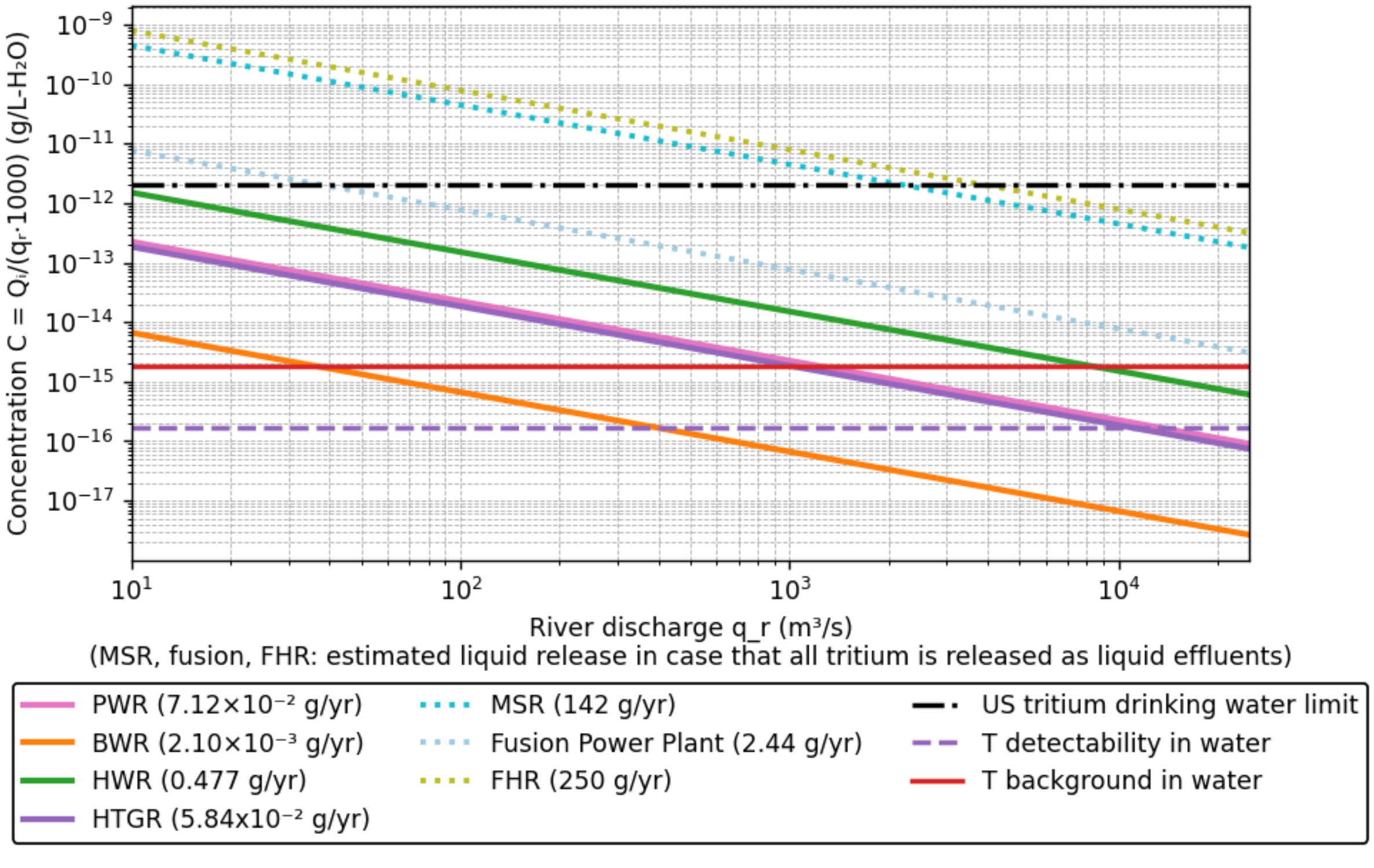

*Figure 5: Resulting hydrologic concentration in a river from liquid effluent releases of tritium from different nuclear facilities using steady-state river mixing (all plants are 1*

*GWe). MSR, FHR, and fusion release curves are estimated in the case that all their tritium is released in liquid form rather than gaseous.*

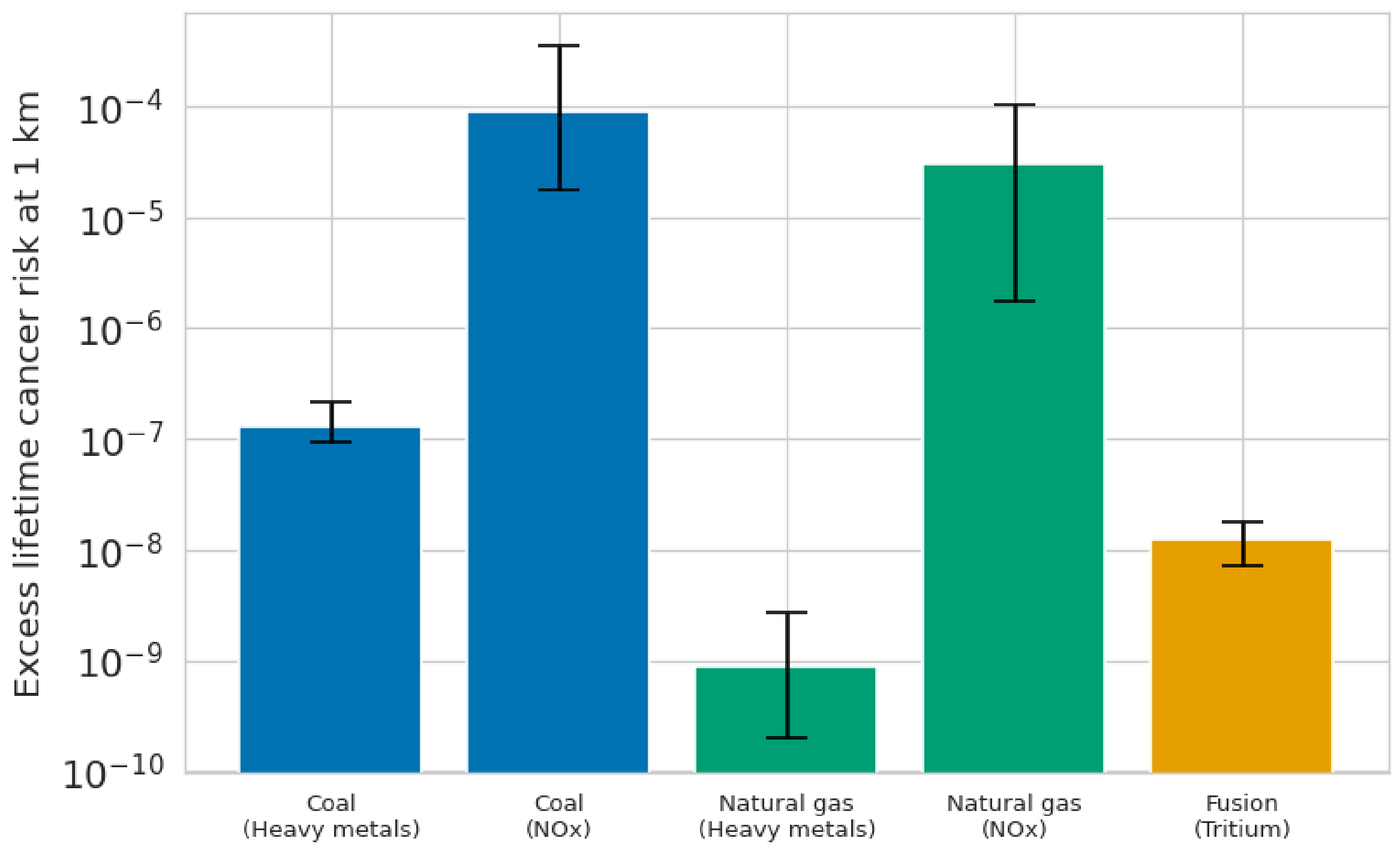

*Figure 6: Risk of excess cancer incidences during normal operation at 1 km from tritium releases from a 1 GWe fusion reactor (depicted in dark yellow), compared to excess risk from chemical pollutant releases (separated by $NO_x$ and heavy metal releases) from a 1 GWe Natural Gas Plant (green) and Brown Coal Plant (blue). Error bars represent the 95% confidence interval for fossil fuel releases, and the highest and lowest release estimates for tritium. The heavy metals included are Ni, As, Cd, and Pb.*

# 6 Data Availability

The datasets generated during and/or analyzed during the current study are available in the GitHub repository, https://github.com/lilianaaaaa/tritium-release.

# 7 Code Availability

All code generated for this paper is available at https://github.com/lilianaaaaa/tritium-release.

# 9 Supplementary Material

| Plant type | PWR | BWR | CANDU | HTGR | Fusion | FHR | MSR |
|---|---|---|---|---|---|---|---|
| **# reactors analyzed** | 62 | 31 | 19 | 1 | | | |
| **# papers data was taken from** | | | | 1 | 2 | 1 | 2 |

*Supplementary Table 1: Summary statistics of the database. Release rate and energy generation values for existing BWR, PWR, and CANDU units are from regulatory and other reports, while reactor release estimates for fusion plants, FHR, and MSR are taken from the literature. Tritium release data for the HTGR was taken from the Fort St. Vrain's*

*operation in the first half of 1986 in addition to estimates from the literature, the values from which are consistent with the data values. Note that CANDU releases represent the summed annual releases from all reactor units at each site.*

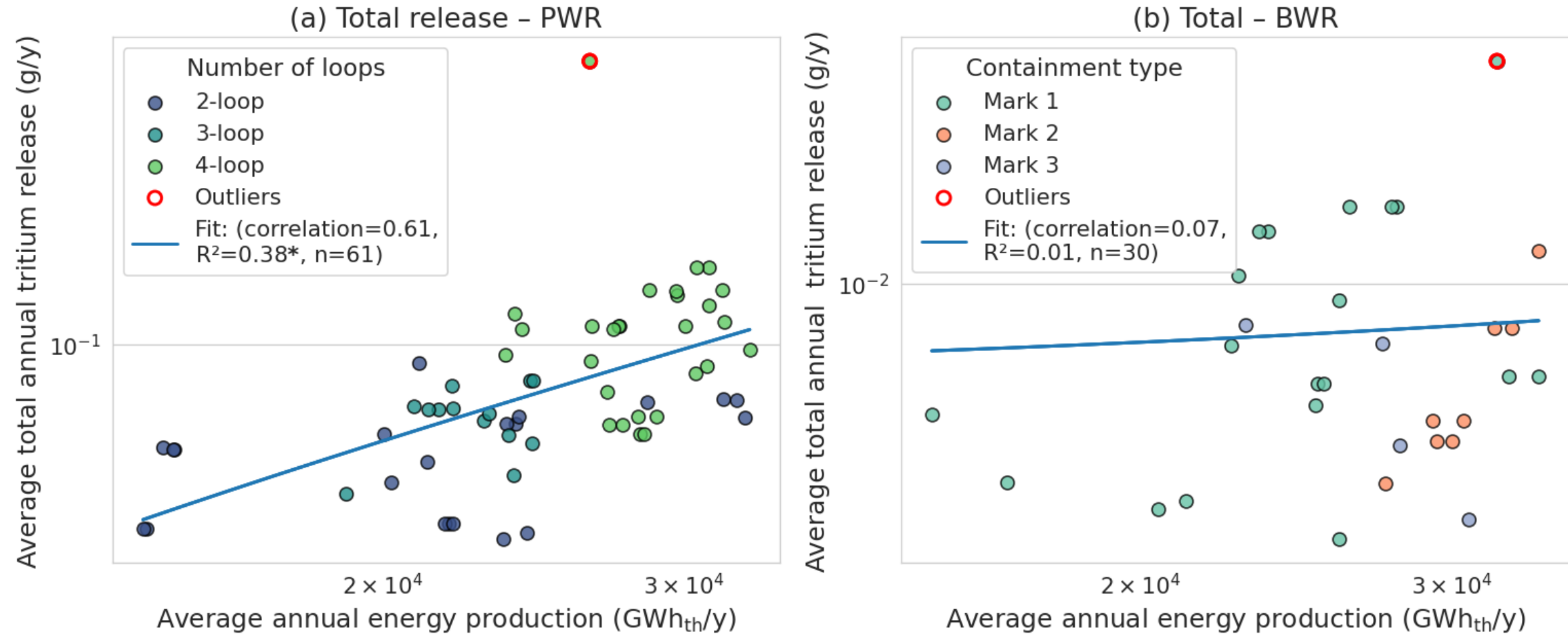

*Supplementary Figure 1: Relationship between thermal annual energy production and total tritium release for LWRs. Different colored points correspond to different numbers of loops for PWRs, and different types of containment for BWRs. Linear regression was conducted to look at the relationship between thermal energy production and total tritium release for both PWRs and BWRs. Outliers were defined as points with standardized residuals larger than 2.5, indicating unusually large deviation from the fitted linear regression. Fitting is based on the inliers. An asterisk next to the R2 value indicates the slope of the regression is statistically significant (p<0.05).*

*Total tritium release from PWRs linearly scales with total thermal energy production, with the slope found to be statistically significant (p<0.05), although there are deviations/scatter. The $R^2$ was found to be 0.38, and the p-value was found to be 1.55 x $10^{-7}$. The only outlier was found to be Watts Bar Unit 1 (tritium is produced for thermonuclear weapons in both Watts Bar Units [1], [2]). Total tritium releases from BWRs were not found to significantly scale with thermal energy production (p > 0.05: p = 7.04 x $10^{-1}$).*

| Pollutant | Value | Unit | 95 percent confidence interval (lower) | 95 percent confidence interval (upper) |
|---|---|---|---|---|
| NOx | 8019154286 | g/GWe-year | 1351542857 | 16669028571 |
| CO | 3514011429 | g/GWe-year | 1802057143 | 5406171429 |
| NMVOC | 234267428.6 | g/GWe-year | 58566857.14 | 937069714.3 |
| SOx (US region) | 25318902.86 | g/GWe-year | 15227382.86 | 35410422.86 |
| SOx (EU region) | 21985097.14 | g/GWe-year | <2703085.714 | 41267108.57 |
| TSP | <12614400 | g/GWe-year | <8109257.143 | <17119542.86 |
| PM10 | <12614400 | g/GWe-year | <8109257.143 | <17119542.86 |
| PM2.5 | <12614400 | g/GWe-year | <8109257.143 | <17119542.86 |
| Pb | <135.1542857 | g/GWe-year | <45.05142857 | <405.4628571 |
| Cd | <22.52571429 | g/GWe-year | <7.208228571 | <67.57714286 |
| Hg | 4505.142857 | g/GWe-year | <126.144 | 90102.85714 |
| As | 10812.34286 | g/GWe-year | <2432.777143 | 32437.02857 |
| Cr | <68.47817143 | g/GWe-year | <22.52571429 | <205.4345143 |
| Cu | <6.847817143 | g/GWe-year | <22.52571429 | <205.4345143 |
| Ni | <45.95245714 | g/GWe-year | <15.31748571 | <137.8573714 |
| Se | <1009.152 | g/GWe-year | <337.8857143 | <3036.466286 |
| Zn | <135.1542857 | g/GWe-year | <45.05142857 | <405.4628571 |

*Supplementary Table 2: Emission factors for public electricity and heat production using natural gas. Emission factors for hard coal, brown coal, and natural gas used for public electricity and heat production were taken from EMEP/EEA [3], and were converted to g/GWe-year (see Methods: Cancer Risk Comparison to Fossil-Energy Systems).*

| Pollutant | Value | Unit | 95 percent confidence interval (lower) | 95 percent confidence interval (upper) |
|---|---|---|---|---|
| NOx | 23967360000 | g/GWe-year | 13875840000 | 55406326154 |
| CO | 844194461.5 | g/GWe-year | 652067446.2 | 5870547692 |
| NMVOC | 135847384.6 | g/GWe-year | 81508430.77 | 326033723.1 |
| SOx | 1.63017E+11 | g/GWe-year | 32021169231 | 4.85169E+11 |
| TSP | 1135296000 | g/GWe-year | 116440615.4 | 11352960000 |
| PM10 | 766567384.6 | g/GWe-year | 97033846.15 | 7665673846 |
| PM2.5 | 310508307.7 | g/GWe-year | 97033846.15 | 3105083077 |
| BC | 3105083.077 | g/GWe-year | 97033.84615 | 124203323.1 |
| Pb | 1455507.692 | g/GWe-year | 1028558.769 | 2396736 |
| Cd | 174660.9231 | g/GWe-year | 125173.6615 | 291101.5385 |
| Hg | 281398.1538 | g/GWe-year | 202800.7385 | 473525.1692 |
| As | 1387584 | g/GWe-year | 999448.6154 | 2338515.692 |
| Cr | 883008 | g/GWe-year | 635571.6923 | 1484617.846 |
| Cu | 97033.84615 | g/GWe-year | 19406.76923 | 485169.2308 |
| Ni | 941228.3077 | g/GWe-year | 685058.9538 | 1601058.462 |
| Se | 4366523.077 | g/GWe-year | 3182710.154 | 7423089.231 |
| Zn | 853897.8462 | g/GWe-year | 48905.05846 | 1630168.615 |
| PCBs | 0.320211692 | g WHO-TEQ/GWe-year | 0.106737231 | 0.960635077 |
| PCDD/F | 0.970338462 | g I-TEQ/GWe-year | 0.485169231 | 1.455507692 |
| Benzo(a)pyrene | 126.144 | g/GWe-year | 25.2288 | 630.72 |
| Benzo(b)fluoranthene | 3590.252308 | g/GWe-year | 359.0252308 | 35902.52308 |
| Benzo(k)fluoranthene | 2813.981538 | g/GWe-year | 281.3981538 | 28139.81538 |
| Indeno(1,2,3-cd)pyrene | 203.7710769 | g/GWe-year | 40.75421538 | 1018.855385 |
| HCB | 650.1267692 | g/GWe-year | 213.4744615 | 1950.380308 |

*Supplementary Table 3: Emission factors for public electricity and heat production using brown coal.*

| Pollutant | Value | Unit | 95 percent confidence interval (lower) | 95 percent confidence interval (upper) |
|---|---|---|---|---|
| NOx | 20280073846 | g/GWe-year | 19406769231 | 33961846154 |
| CO | 844194461.5 | g/GWe-year | 596758153.8 | 1455507692 |
| NMVOC | 97033846.15 | g/GWe-year | 58220307.69 | 232881230.8 |
| SOx | 79567753846 | g/GWe-year | 32021169231 | 4.85169E+11 |
| TSP | 1106185846 | g/GWe-year | 291101538.5 | 29110153846 |
| PM10 | 747160615.4 | g/GWe-year | 194067692.3 | 19406769231 |
| PM2.5 | 329915076.9 | g/GWe-year | 87330461.54 | 8733046154 |
| BC | 7258131.692 | g/GWe-year | 235792.2462 | 705630129.2 |
| Pb | 708347.0769 | g/GWe-year | 500694.6462 | 1164406.154 |
| Cd | 87330.46154 | g/GWe-year | 60840.22154 | 141669.4154 |
| Hg | 135847.3846 | g/GWe-year | 98974.52308 | 230940.5538 |
| As | 688940.3077 | g/GWe-year | 489050.5846 | 1144999.385 |
| Cr | 436652.3077 | g/GWe-year | 310508.3077 | 723872.4923 |
| Cu | 756864 | g/GWe-year | 22608.88615 | 1504024.615 |
| Ni | 475465.8462 | g/GWe-year | 333796.4308 | 779181.7846 |
| Se | 2231778.462 | g/GWe-year | 1552541.538 | 3619362.462 |
| Zn | 1843643.077 | g/GWe-year | 752012.3077 | 15040246.15 |
| PCB | 0.320211692 | g WHO-TEQ/GWe-year | 0.106737231 | 0.960635077 |
| PCDD/F | 0.970338462 | g I-TEQ/GWe-year | 0.485169231 | 1.455507692 |
| Benzo(a)pyrene | 67.92369231 | g/GWe-year | 23.77329231 | 214.4448 |
| Benzo(b)fluoranthene | 3590.252308 | g/GWe-year | 359.0252308 | 35902.52308 |
| Benzo(k)fluoranthene | 2813.981538 | g/GWe-year | 281.3981538 | 28139.81538 |
| Indeno(1,2,3-cd)pyrene | 106.7372308 | g/GWe-year | 57.34700308 | 228.9998769 |
| HCB | 650.1267692 | g/GWe-year | 213.4744615 | 1950.380308 |

*Supplementary Table 4: Emission factors for public electricity and heat production using hard coal.*

| **Tritium mitigation strategy** | **Description** | **Reference** |
|---|---|---|
| Permeation window | Hydrogen isotopes are removed from gas mixtures as they permeate through walls of permeable material at a metal window (Fukada and Mitsuishi [4]). | Lam et al. [5]<br>Fukada and Mitsuishi [4] |
| Stripping column | Tritium dissolved in a liquid enters into the column from the top in hydrogen saturated conditions. Gas is injected in the column from the bottom so that the gas bubbles are fragmented. The gas passes through the liquid | Lam et al. [5]<br>Aiello et al. [6] |
| | and strips the liquid of hydrogen, and it is then collected by a dome (Aiello et al. [6]) | |
| Fixed graphite adsorption column | The column is modeled in the hot leg of the system, where the graphite in the column adsorbs tritium (Lam et al. [5]). | Lam et al. [5] |
| Vapor Phase Catalytic<br>Exchange (VPCE) followed by Cryogenic Distillation | Firstly, water is evaporated and mixed with cold deuterium gas. The mixture is superheated to 200°C and enters the catalyst bed, where tritium is exchanged and equilibrated. The deuterium gas is separated from water in the condenser and returned to the cryogenic distillation system. The water, now tritium-depleted, goes to the evaporator of the next stage. VPCE process typically consists of three to eight stages, with each stage having a water evaporator, superheater, catalyst bed and condenser– separator (IAEA [7]). | Electric Power Research Institute [8]<br>Busigin et al. [9]<br>IAEA [7] |
| Combined Electrolysis Catalytic Exchange (CECE) | This process is based on the hydrogen/water exchange equilibrium reaction:<br><br>$HT\ (g) + H_20\ (l) \leftrightarrow HTO\ (l) + H_2\ (g)$<br><br>which favors HTO formation when $H_20$ (l) is in contact with HT (g) (Electric Power Research Institute [8]).<br><br>CECE consists of countercurrent gas/liquid exchange columns with packed catalyst beds, an electrolysis cell and a hydrogen/oxygen recombiner (Electric Power Research Institute [8]). The incoming water stream is added midcolumn, and as it flows down, tritium is transferred from a HT (g) stream, producing tritiated water. $H_20$ (l) is added to dilute to acceptable release levels. Additionally, the liquid is electrolytically split into oxygen and HT (g). | Electric Power Research Institute [8]<br>Busigin et al. [9]<br>IAEA [7] |
| Molecular separation | HTO is selectively adsorbed through hydration of ions loaded on a conditioned resin media, | Electric Power Research Institute [8] |

| | and the captured tritium is released from the conditioned resin media through a stepwise drying process (Electric Power Research Institute [8]). | |
|---|---|---|
| Bithermal Hydrogen-Water Processing | This process is based on the same reaction as CECE and uses the same catalysts, but relies on a recycled stream of hydrogen coupled with dual temperature separation columns (Electric Power Research Institute [8]). This process consists of cold and hot enriching and stripping columns in vertical orientation, where hydrogen gas flows upwards and the liquid stream flows downwards. The upper cold stripper strips tritium from the hydrogen, and tritium-free gas recirculates to the hot stripper column to remove tritium from the wastewater. The tritiated stream is separately captured for disposal. | Electric Power Research Institute [8] |
| Intermediate nitrate-salt loop | Used for secondary tritium trapping (Forsberg [10]). Because hot nitrate salts are highly oxidizing, they create oxide layers that have low permeability to tritium diffusion. The oxidizing salts oxidize any tritium collected in the offgas system to steam, which does not diffuse through steel (Forsberg et al. [11]). | Forsberg [10]<br>Forsberg et al. [11] |

*Supplementary Table 5: Mitigation strategies for tritium release.*

## 9.1 Supplementary References

# 10 Acknowledgements

We would like to acknowledge Charles Forsberg for providing information on tritium leakage rate in reactors using molten salt. We would also like to thank Dennis Whyte and George Tynan for providing feedback. This material is based upon work supported by the Department of Energy National Nuclear Security Administration through Defense Nuclear Nonproliferation's Enabling Capabilities in Technology Consortium under Award Number DE-NA0004197. This report was prepared as an account of work sponsored by an agency of the United States Government. Neither the United States Government nor any agency thereof, nor any of their employees, makes any warranty, express or limited, or assumes any legal liability or responsibility for the accuracy, completeness, or usefulness of any information, apparatus, product or process disclosed, or represents that its use would not infringe privately owned rights. Reference herein to any specific commercial product, process, or service by trade name, trademark, manufacturer, or otherwise does not necessarily constitute or imply its endorsement, recommendation, or favoring by the United States Government or any agency thereof. The views and opinions of authors herein do not necessarily state or reflect those of the United States Government or any agency thereof. LLNL co-author contributions were performed under Contract DE-AC52-07NA27344. LLNL-JRNL-XXXXXX.

## 11 Author Contributions

L.A. and H.M.W. worked on the conceptualization, methodology and analysis, in addition to the writing of the first draft. J.W. supported the calculation and analysis of the Gaussian Plume and cancer risk. A.V. supported the tritium release and inventory data analysis. H.K., S.U. and N.A. provided relevant information and texts for environmental tritium concentrations and measurement techniques All the authors reviewed and edited the draft.

## 12 Competing Interests

The authors declare no competing interests.